\documentstyle[epsfig]{article}
	\newlength{\paperbaselineskip}
\newfont{\fourteencp}{cmcsc10 scaled\magstep2}
\newfont{\titlefont}{cmbx10 scaled\magstep2}
\newfont{\authorfont}{cmcsc10 scaled\magstep1}
\newfont{\fourteenmib}{cmmib10 scaled\magstep2}
	\skewchar\fourteenmib='177
\newfont{\elevenmib}{cmmib10 scaled\magstephalf}
	\skewchar\elevenmib='177
\newfont{\ninemib}{cmmib9} \skewchar\ninemib='177
\makeatletter
\newif\ifpr@pstyle \pr@pstylefalse
\newif\ifnons@qeq  \nons@qeqfalse
\newcommand\nonsequentialeqnum{
        \nons@qeqtrue
	\@addtoreset{equation}{section}
	\def\theequation{\arabic{section}.\arabic{equation}}}
\newif\ifp@bblock  \p@bblocktrue
\newcommand\nopubblock{\p@bblockfalse}
\newcommand\topspace{\hrule height 0pt depth 0pt \vskip}
\newcommand\p@bblock{\begingroup \tabskip=\hsize minus \hsize
	\baselineskip=1.5\ht\strutbox \topspace-2\baselineskip
	\halign to\hsize{\strut ##\hfil\tabskip=0pt\crcr
	\the\Pubnum\crcr\the\date\crcr}\endgroup}
\newcommand\YUKAWAmark{\hbox{
        \ifpr@pstyle\ninemib\else\elevenmib\fi
        Yukawa\hskip1mm Institute\hskip1mm Kyoto \hfill}}
\newtoks\date
\newtoks\Pubnum

\newcommand{\frontpageskip}{\vspace{12pt plus .5fil minus 2pt}}
\def\@authoraddress{} \def\@title{}
\def\title#1{\gdef\@title{\frontpageskip
	\begin{center}{\titlefont #1}\end{center}\par}}
\def\@author#1{\frontpageskip\par\begin{center}{\authorfont #1}
	\end{center}
	\nobreak}
\def\author#1{\expandafter\def\expandafter\@authoraddress\expandafter
    {\@authoraddress{\@author{#1}}}}
\def\andauthor#1{\expandafter\def\expandafter\@authoraddress\expandafter
    {\@authoraddress{\frontpageskip\centerline{and}\@author{#1}}}}
\def\authors#1{\expandafter\def\expandafter\@authoraddress\expandafter
    {\@authoraddress{\frontpageskip\noindent #1}}}
\def\@address#1{\par\begin{center}{\sl #1}\end{center}\par}
\def\address#1{\expandafter\def\expandafter\@authoraddress\expandafter
    {\@authoraddress{\@address{#1}}}}
\def\andaddress#1{\expandafter\def\expandafter%
    \@authoraddress\expandafter
    {\@authoraddress{\par\centerline{\sl and}\@address{#1}}}}
\renewcommand{\thanks}[1]{\footnote{#1}}
\def\maketitle{\par
  \begingroup
       \def\thefootnote{\fnsymbol{footnote}}
	\thispagestyle{empty}
        \baselineskip=\paperbaselineskip
	\@maketitle
	\endgroup
	\setcounter{footnote}{0}
	\let\maketitle\relax \let\@maketitle\relax
	\let\@thanks\relax \let\@title\relax
	\let\@title\relax \let\@authoraddress\relax
	\let\thanks\relax}
\def\@maketitle{%
        \ifpr@pstyle\vspace{-1.0cm}\else\vspace{-1.7cm}\fi
	\YUKAWAmark\vskip0.6cm
	\ifp@bblock\p@bblock \else\hrule height 0pt \relax \fi
	\@title
	\@authoraddress
	}
\renewcommand{\abstract}{\par\vspace{80pt plus 3pt minus 3pt}
	\frontpageskip\centerline{
             \ifpr@pstyle\twelvecp\else\fourteencp\fi Abstract}
	\vspace{8pt plus 3pt minus 3pt}}
\makeatother
\def\comptblwd#1#2{%
\expandafter\gdef\csname tref@#1\endcsname{#2}%
}
\newcommand{\Ge}{$^{64}$Ge}
\newcommand{\Se}{$^{68}$Se}
\newcommand{\Kr}{$^{72}$Kr}
\newcommand{\Sr}{$^{76}$Sr}
\newcommand{\Zr}{$^{80}$Zr}
\newcommand{\Yzero}{$Y_{30}$}
\newcommand{\Yone}{$Y_{31}$}
\newcommand{\Ytwo}{$Y_{32}$}
\newcommand{\Ythree}{$Y_{33}$}
\newcommand{\Td}{$T_d$}

\begin{document}
\Pubnum={YITP-98-69}
\date={October 1998}

\title{\LARGE\bf\boldmath Non-axial Octupole Deformations of 
$N=Z$ Nuclei \\ 
in  $A \sim 60-80$ Mass Region \thanks{A talk presented by M.M. at
{\it Nuclear Structure '98}, Gatlinburg, August 10-15, 1998}}

\author{M.~Matsuo}
\address{
Yukawa Institute for Theoretical Physics, Kyoto University,
Kyoto 606-8502, Japan}
\author{S.~Takami  and K.~Yabana}
\address{
Graduate School of Science and Technology,
Niigata University, Niigata 950-2101, Japan}

\maketitle

\begin{abstract}
By performing a fully three dimensional Hartree-Fock 
calculation with use of the Skyrm forces, we demonstrate 
possibility of exotic deformations violating both 
the reflection and the axial symmetries of $N=Z$ nuclei in 
$A \sim 60-80$ mass region. 
The \Ytwo\ tetrahedral shape predicted in excited \Zr\ arises from
a shell gap at $N,Z = 40$ which is enhanced for the tetrahedron
deformation. Softness toward the \Ythree\ triangular deformation 
of the oblate state in \Se\ is also predicted.
\vfill
\end{abstract}

\section{Introduction}
Octupole deformations that violate the reflection symmetry
have attracted many attentions
recently in studies of nuclear structure\cite{BN96}.
Although axial octupole deformation is well established 
in actinides and in neutron rich Xe and Ba region, 
there seems no experimental evidence of more
exotic shapes that violate both the reflection and the axial
symmetries, 
except for light alpha-clustering nuclei such as
$^{12}$C \cite{Fujiwara}.

Non-axial octupole deformations have been predicted theoretically for
actinides \cite{LD94} and in superdeformed nuclei \cite{Li-SD,Chasman}.
Instead we point out possibility of the non-axial octupole shapes in
$N=Z$ $A\sim 60-80$ region of the nuclear chart \cite{Takami98,Skalski}.
For $N=Z$ nuclei 
the deformation driving shell effects which arise from both  
protons and neutrons cooperate coherently.
The shell effect is indeed large in this mass region
as is illustrated by the prolate-oblate shape coexistence
in Kr and Se isotopes \cite{shapecoex} and the sudden onset of large prolate
deformation for $N,Z \approx 38-40$ \cite{NZnuclei}.
Furthermore octupole instability due to the single-particle  structure
 in nuclear mean-field  is predicted 
for $N,Z \sim 34$ (in addition to $56,90,134$) \cite{Nea84,Lea82} for which
the coupling between $g_{9/2}$ and  $p_{3/2}$ orbits are
mostly responsible. 

It is found recently that the shell structures plays
important roles for aggregates of metallic atoms (metal clusters).
In this system the mean-field for the valence electrons resembles
with the nuclear mean-field except for its effectively zero
$ls$ potential \cite{mcluster}. Non-axial octupole deformatios
seem to be one of the dominant deformations in this system 
as demonstrated by  means of schematic models 
\cite{Hamamoto,Frisk} and more recent
mean-field calculations \cite{Reimann,Kole}. 
In particular, the tetrahedron shape
or the \Ytwo\ deformation is predicted to be stable because of 
the shell gaps generated at $N_{electron}=40,70,112,156$ 
\cite{Hamamoto,Reimann,Kole}. 
In this paper, we will show that
the tetrahedron shape is expected in the nucleus \Zr\ with $N=Z=40$,
being in parallel with the metal cluster system in spite of
the difference in the $ls$ potential and the pairing correlation
present in nuclei. Furthermore, non-axial octupole
deformations besides \Ytwo\ are important deformation modes in 
$N=Z$ nuclei 
in $A \sim 60-80$ mass region.  See also 
the previous publication \cite{Takami98}.

\section{A fully three dimensional Hartree-Fock calculation}

For investigating non-axial reflection asymmetric
deformations by means of the mean-field theory, 
we have to exclude any symmetry assumptions 
on nuclear deformations. Furthermore, it is important
to use a description which allows arbitrary deformation when we 
search unknown shapes. From this view point,
we adopt the Hartree-Fock method and perform a fully three
dimensional calculation using the Cartesian mesh
representation 
without any requirements on nuclear shapes and its
symmetries. The Skyrm force is used as the effective nucleon-nucleon
force. The pairing correlation is treated by means of the
BCS method for the seniority force with a cutoff of the
single-particle orbits, which is chosen the same as Ref.
\cite{Bonche85,Tajima}. The imaginary time method is used
for the iteration procedure \cite{Davies}. 
Thus the basic formulation is in parallel with
that of Ref.\cite{Bonche85}, except for the symmetry treatment 
(they assumed the reflection symmetries) and calculational
details. A new code has been developed independently so that it 
can also include the parity projection and full variation after
projection\cite{TakamiPHF}.
The
parity projected Hartree-Fock with Skyrm force, but without BCS
treatment, has been applied to light nuclei\cite{TakamiPHF}
(but not to the present investigation).

For the results presented below, we use the 3D mesh which are
enclosed by a sphere of radius $R=13$ fm, and the mesh width of
1 fm. The parameter set of SIII \cite{SIII} is used for the Skyrm
force in most calculations while SkM$^*$ \cite{SkMS} 
is also employed for comparison.
Concerning the pairing force, we use  
$G_{p}=16.5/(11+Z)$ MeV for protons as given in Ref. \cite{Bonche85}.
For neutrons, we use $G_{n}=16.5/(11+N)$ MeV  \cite{Heenen} since
it is natural to take the same value as
protons for $N=Z$ nuclei.
In order to search not only the ground state but also 
local minimum states, we performed different
runs of iterations 
starting with different initial conditions.
To characterize deformation of the obtained solutions, we use 
the mass multipole moments,
\begin{equation}
	\alpha_{lm} \equiv \frac{4\pi \langle \Phi | 
		\sum_{i}^{A} r_{i}^{l} X_{lm}(i) | \Phi \rangle}
		{3 A R^{l}}, (m=-l,\cdots,l),
\end{equation}
where $A$ is the mass number and $R = 1.2 A^{1/3}$ fm.
Here $X_{lm}$ is a  real basis of the spherical harmonics,
\begin{eqnarray}
	X_{l0} & = & Y_{l0},
		\nonumber \\
	X_{l|m|} & = & \frac{1}{\sqrt{2}}( Y_{l-|m|}+Y_{l-|m|}^{*} ),
		\nonumber \\
	X_{l-|m|} & = & \frac{-i}{\sqrt{2}} ( Y_{l|m|}-Y_{l|m|}^{*} ),
\end{eqnarray}
where the quantization axis is chosen as  the largest and smallest
principal inertia axes for  prolate and oblate solutions, respectively.
We put the constraints $\alpha_{1m}=0 (m=-1,0,1)$  for the center of mass,
and $\alpha_{2m}=0 (m=-2,-1,1)$ for the principal axes. This is done
by adding the constraining fields $-\lambda_{lm}r^lX_{lm}$
to the mean field, where the
constraining Lagrange multipliers are determined 
so that the constraints are satisfied in each iteration step\cite{TakamiPHF}.

\begin{table*}[b!]
\caption{The ground and local minimum HF solutions obtained with the SIII
        force. The number in the first line is 
the excitation energy in MeV measured
from the ground state solution. The second and the third line
list the quadrupole and octupole deformations of the solutions.}
\begin{tabular}{c||cccc}
\hline
         & {Oblate $\beta>0,\ 30^{\circ}<\gamma\le60^{\circ}$} 
         &  $\beta=0$ 
         & \multicolumn{2}{c}{Prolate $\beta>0, \ 
0^{\circ}\le\gamma\le30^{\circ}$} \\ 
           \hline \hline

          &   &     & g.s. &  \\
 $^{64}$Ge
         & 
         & 
         & $\beta ,\gamma = 0.28, 25^{\circ}$ (triaxial) 
         & \\
         & 
         &
         &  $\beta_{3} = 0.00$
         &    \\ \hline

         &  g.s. &   & 0.25 &  \\

 $^{68}$Se
         & $\beta ,\gamma  = 0.28, 60^{\circ}$ 
         & 
         & $\beta ,\gamma = 0.25, 0^{\circ}$ 
         & \\

         &  $\beta_{3} = \beta_{33}=0.14$ 
         &  
         &  $\beta_{3} = 0.00$ 
         &     \\ \hline

         & g.s.   & & 1.47 & \\
 $^{72}$Kr
         & $\beta ,\gamma = 0.35, 60^{\circ}$
         & 
         & $\beta ,\gamma = 0.40, 0^{\circ}$ 
         & \\

         &  $\beta_{3} = 0.00$
         &  
         &  $\beta_{3} = 0.00$ 
         &    \\ \hline

         &  2.72 &  & & g.s. \\
 $^{76}$Sr
         & $\beta ,\gamma  = 0.14, 60^{\circ}$ 
         & 
         &  
         & $\beta ,\gamma = 0.50, 0^{\circ}$ \\

         &  $\beta_{3}=0.13,\beta_{32}=0.12$
         & 
         &  
         &  $\beta_{3} = 0.00$ \\ 
\hline

         &  1.56 & 0.96 & & g.s. \\
 $^{80}$Zr
         & $\beta ,\gamma  = 0.19, 60^{\circ}$ 
         & $\beta ,\gamma = 0.00, 0^{\circ}$
         & 
         & $\beta ,\gamma = 0.51, 0^{\circ}$ \\
 
         &  $\beta_{3} =0.00$ 
         &  $\beta_{3} = \beta_{32}=0.23$
         &  
         &  $\beta_{3} = 0.00$ \\ 
\end{tabular}
\end{table*}

For the quadrupole moment, we  use ordinary
($\beta,\gamma$) notation, i.e., $\alpha_{20}=\beta\cos{\gamma},
\alpha_{22}=\beta\sin{\gamma}$, mapped in the $\beta>0, 0<\gamma<
\pi/3$ section. To represent magnitude of the octupole deformation, we define
\begin{equation}
        \beta_{3}\equiv ( \sum_{m=-3}^{3} \alpha_{3m}^{2} )^{\frac{1}{2}},
        \; \; \; \; \; \;
        \beta_{3m} \equiv ( \alpha_{3m}^{2}+\alpha_{3-m}^{2} )^{\frac{1}{2}}
        \; \; \; \; \; \;
          (m=0,1,2,3).
\end{equation}

\section{Shapes of ground and local minimum states}

The HF ground states and local minimum solutions 
for the even-even nuclei \Ge, \Se, \Kr, \Sr\ and \Zr\ are 
listed in Table 1.  
The result is essentially the same as the previous
calculation \cite{Takami98} while  the convergence
of the HF iteration is taken care of more carefully in
the present calculation. It is noted that many of these
nuclei have local minimum solutions with low excitation
energy, suggesting presence of shape coexisting excited
states. The ground state shape changes 
from the triaxial (\Ge),  the oblate (\Se,\Kr), 
to the strong prolate shape (\Sr,\Zr) as $N,Z$ increases.
This is  consistent with the experimental trends
\cite{NZnuclei,GeEnnis,SeSkoda}.

The reflection asymmetric solutions
with non-zero octupole deformation ($\beta_3 >0$) are obtained
as the ground state of \Se, and as the shape coexisting excited 
states in \Sr\ and \Zr. These solutions {\it violate not only
the reflection symmetry but also the axial one}. 
The \Se\ ground state has finte \Ythree\ deformation ($\beta_{33}=0.14$)
together with oblate quadrupole deformation. The density profile
shows a triangular shape, as seen in Fig.1.
In \Zr, the first local minimum solution with the
excitation energy of 0.96 MeV has large \Ytwo\ deformation
($\beta_{32}=0.23$) whereas the solution has no quadrupole deformation.
It has a tetrahedral shape as shown in Fig. 1. Although this
solution violates both the reflection and the axial symmetry,
the tetrahedral symmetry (\Td) of the point group emerges
as a spontaneous symmetry. Indeed the calculated single-particle
energy spectrum shows a four-fold degeneracy which is a characteristic
feature of the Fermion mean-field with the \Td\ symmetry \cite{LD94}
(See also Fig.3(c)).

\begin{figure}[h!] 
\begin{minipage}{75mm}
\epsfig{file=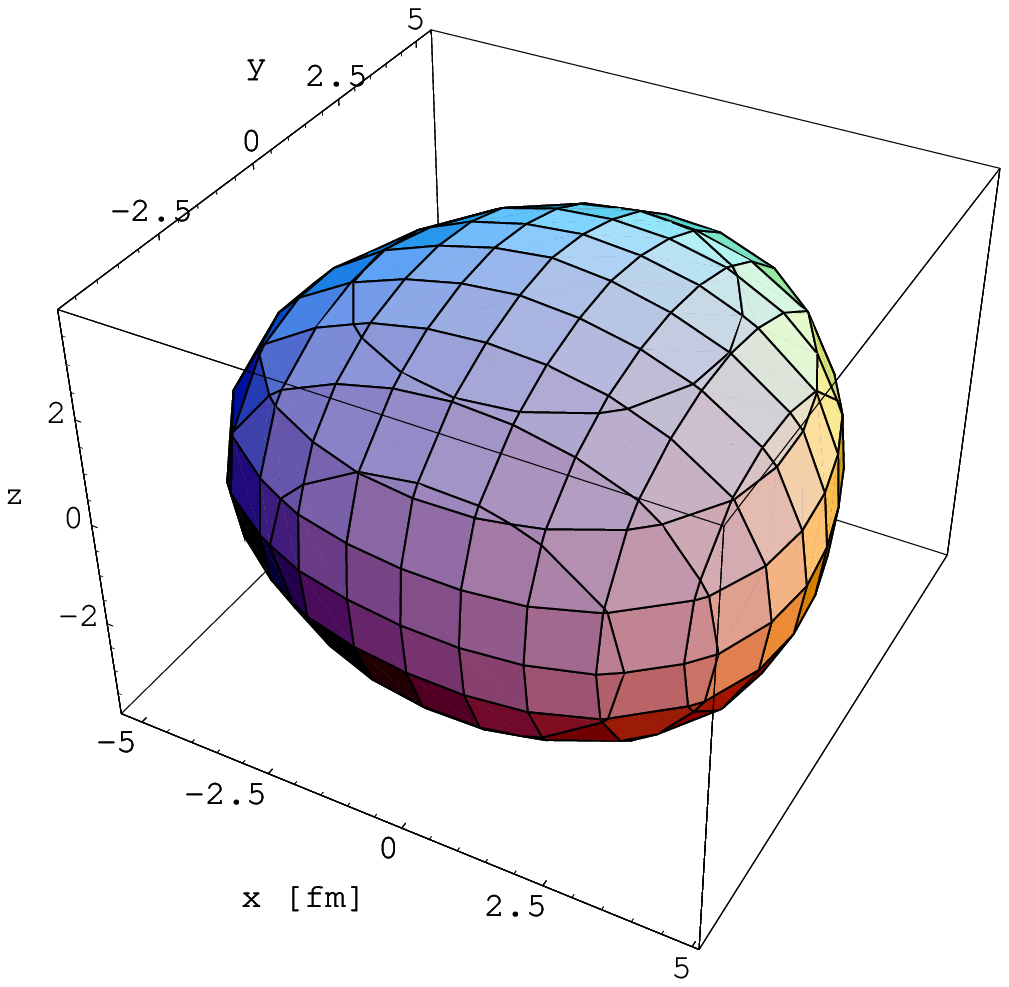,width=75mm}
\end{minipage}
\hspace{\fill}
\begin{minipage}{80mm}
\epsfig{file=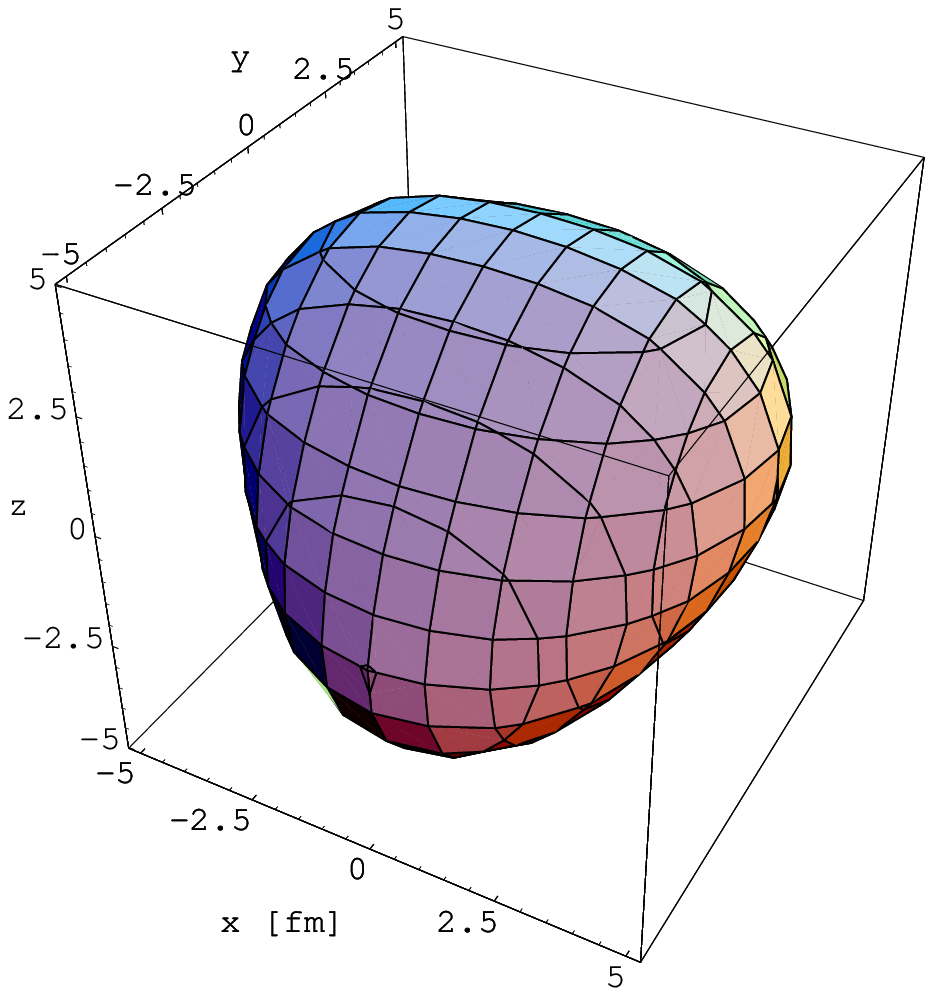,width=79mm}
\end{minipage}
\vspace*{10pt}
\caption{The density contour surface at the
half central density for the triangular ground state solution in 
\Se\ (left), and the
tetrahedral first local minmum solution in \Zr\  (right), listed
in Table 1 obtained with the  SIII force.
}
\label{fig1}
\end{figure}

\section{Non-axial octupole deformations}

\begin{figure}[p] 
\centerline{
\hspace{40mm}
\epsfig{file=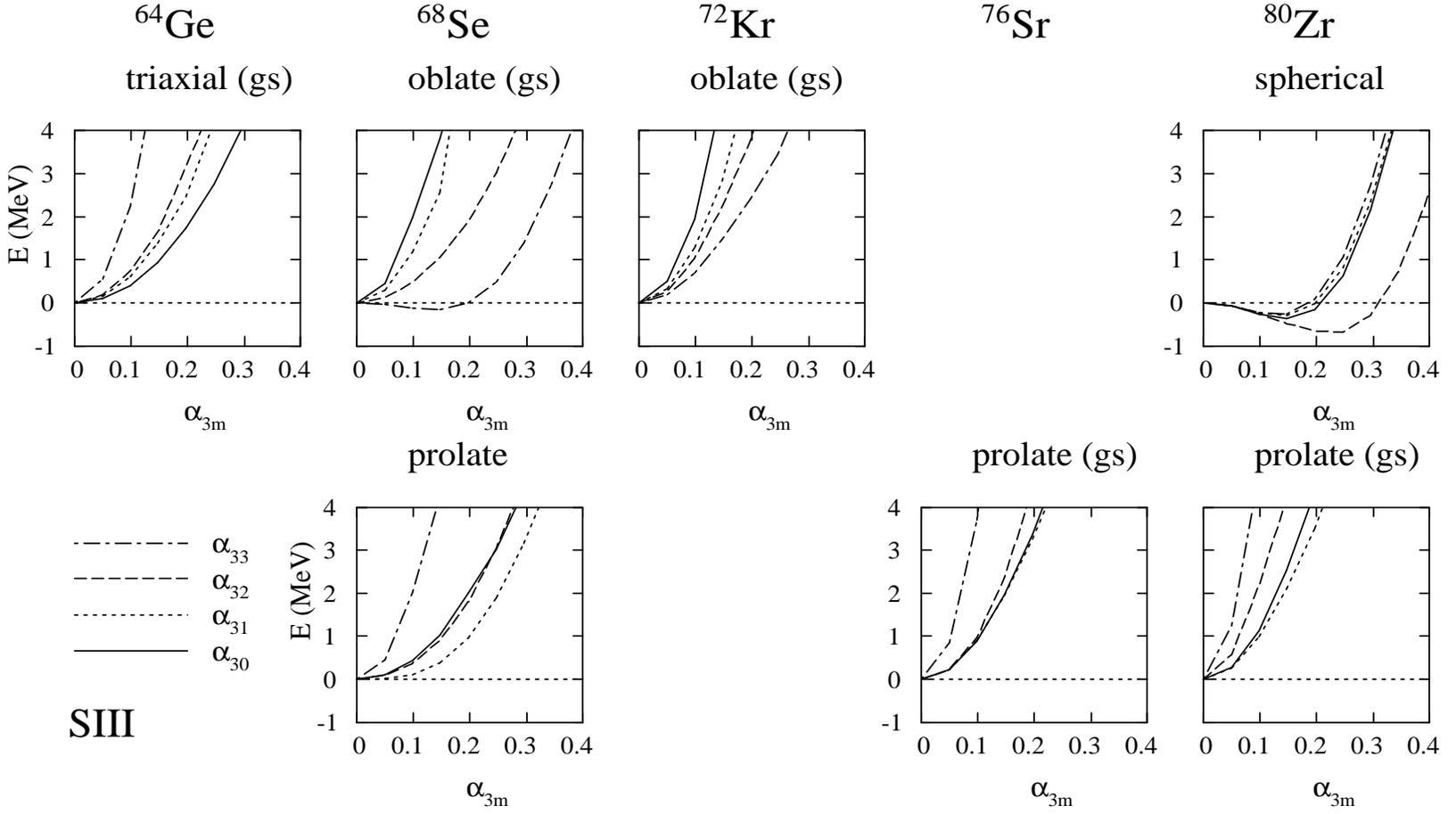,width=12cm}}
\vspace*{5pt}
\caption{Potential energy curve as a function of the octupole
deformation $\alpha_{3m} (m=0,1,2,3)$ associated with the ground
and local minimum solutions listed in Table 1.}
\label{fig2}
\end{figure}

\subsection*{Potential energy curves}

To evaluate softness of the obtained solutions toward the octupole
deformations, we calculate the potential energy curve as
a function of the deformation parameters $\alpha_{3m}$
for all the independent components $m=0,1,2,3$ 
of octupole deformations 
by means of the constraint Hartree-Fock calculation. For this calculation,
we introduce additional nine constraints for the quadrupole
deformation $\alpha_{20},\alpha_{22}$
(equivalent to $\beta,\gamma$) and the octupole deformations 
$\alpha_{3m} (m=-3,..,3)$.
The values of $\beta,\gamma$ is fixed to the ones of a HF solution
under consideration,
and one of $\alpha_{3m} (m=0,1,2,3) $ is varied while other
 $\alpha_{3m}$'s are fixed to zero.
The calculated potential energy curve is plotted in Fig.2.

\subsection*{Tetrahedron shape and associated shell gap in \Zr}

\begin{figure}[b!] 
\centerline{\hspace{30mm}\epsfig{file=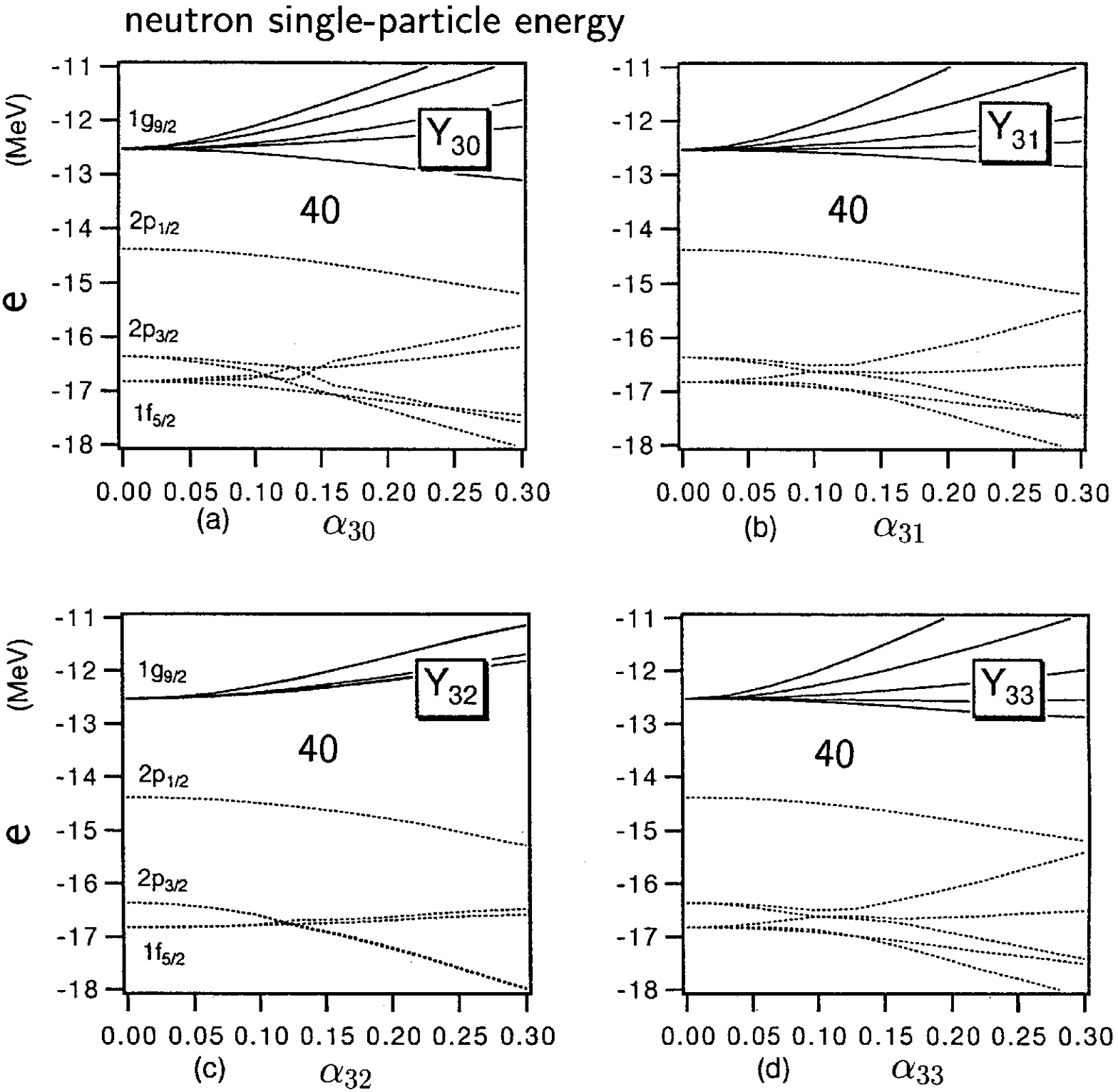,width=13.5cm}}
\vspace*{10pt}
\caption{Neutron single-particle energies 
as a function of octupole deformation $\alpha_{3m}$ for $m=0,1,2,3$
((a),(b),(c),(d), respectively) calculated for \Zr\ with $\beta=0$.
The quadrupole constraint is fixed to $\beta=0$. The SIII force is
used. The proton spectrum is almost the same as neutrons.
}
\label{fig3}
\end{figure}

The tetrahedral solution in \Zr\ corresponds to the minimum
at $\alpha_{32}=0.23$ of the $\alpha_{32}$ potential energy curve 
with the spherical constraint ($\beta=0$).
The energy gain due to the tetrahedron deformation or
equivalently the energy difference between the minimum at
$\alpha_{32}=0.23$ relative to $\alpha_{32}=0$
is as large as 0.70 MeV. 
It is noted that the potential
energy curves for  $\alpha_{3m} (m=0,1,3)$ do not show such deep
minimum. The reason of this can be seen in the single-particle
energy diagram, Fig.3, plotted as a function of the octupole
deformations $\alpha_{3m}$. Figure 3 indicates that the single-particle
energy spectrum accompanies a large gap at $N,Z=40$ which increases as
increasing the tetrahedral deformation $\alpha_{32}$ 
while the other octupole deformations $m=0,1,3$ 
do not have this feature. 
This is the gap specific to the tetrahedron deformation, and presence
of such gap is due to the high symmetry (the \Td\ symmetry of the
point group) \cite{Hamamoto}.
Note that the tetrahedron shell gap at particle number 40 is found
also in the metal cluster systems \cite{Kole} for which there is
no $ls$ potential. Figure 3 indicates that the tetrahedron shell gap
emerges even under influence of the large $ls$ term in nuclear potential.
It is also noted that the tetrahedron minima is obtained with including
the pairing. The realistic
Hartree-Fock calculation thus supports the correspondence between 
nuclei and clusters suggested in Ref.\cite{Kole} on the presence
of the tetrahedron shell gap.

\subsection*{Triangular Softness in \Se}

Concerning the oblate triangular solution of \Se, the minima is
quite shallow with respect to the $\alpha_{33}$ direction.
The energy gain due to the \Ythree\ deformation (the energy
difference between the minimum and the solution with 
$\beta_3,\alpha_{33}=0$ ) is just 0.15 MeV. This indicates that
the oblate state is extremely soft toward the  $\alpha_{33}$
deformation rather than a rigid triangular deformation. 
It is noted that the potential energy curves  for $\alpha_{3m}=0 (m=0,1,2)$ 
are not as soft as the $\alpha_{33}$ curve.

\subsection*{Systematics}

In Fig.2 and Table 1 are seen systematic trends of the non-axial octupole
deformations in $N=Z, A\sim 60-80$ nuclei. The octupole softness
is enhanced  at $N,Z\sim 34$, (representative is oblate ground state of \Se\
which is soft for triangular deformation),
and at $N,Z \sim 40$ (the tetrahedral deformation). It is also seen
that the \Ythree\ mode is the softest among the four octupole deformations
for the oblate states whereas \Yone\ and \Yzero\
is favored for the prolate states.
This is explained in terms of the single-particle shell structures
near the Fermi surface\cite{Takami98}. For the oblate states, 
the high $\Omega$  orbits such as [404]9/2 and [413]7/2
stemming from $g_{9/2}$ are located near the Fermi surface of $N,Z\sim 34$
and strong \Ythree\  coupling with [301]3/2 and [310]1/2 orbits emerges.
On the other hand the $g_{9/2}$ low $\Omega$ orbits are 
situated far above the Fermi surface.  Thus 
\Yzero\ \Yone\ couplings are  disfavored.
For the prolate states  \Yone\ is favored in a reversed reasoning.
The \Ytwo\ deformations favored at $N,Z\sim 40$ is caused by the tetrahedron
shell gap discussed above.
The instability and the softness for the non-axial
octupole deformations are quite contrasting with the axial octupole
instability known in other mass regions.

\section{Discussion}

\begin{figure}[t] 
\centerline{\epsfig{file=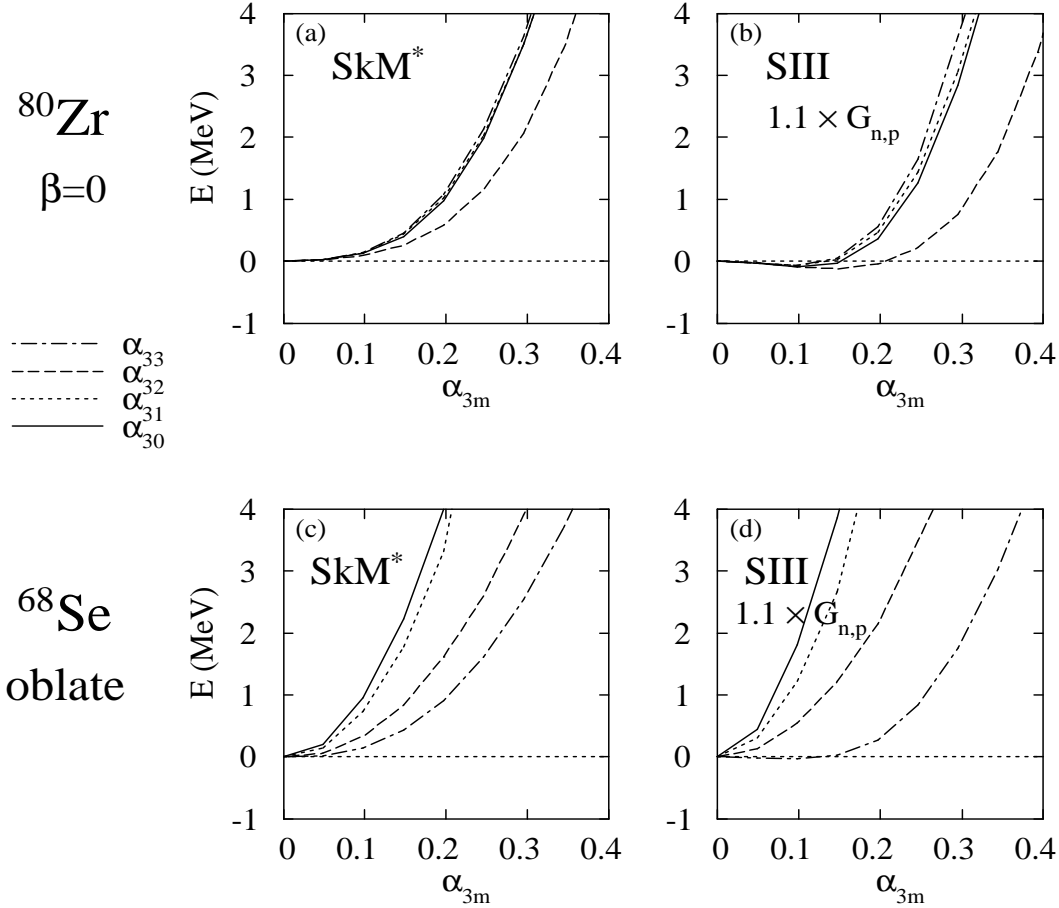,width=12cm,angle=-90}}
\vspace*{10pt}
\caption{Potential energy curve calculated with the SkM$^*$ force
(a,c), and with SIII and an increased pairing force $G_{n,p}=18.15/(11+N,Z)$
MeV (b,c), for the excited spherical state in \Zr\ (a,b)
and the oblate state in \Se\ (c,d).
}
\label{fig4}
\end{figure}

\subsection*{Sensitivity to force parameters}

Besides SIII, other parameter sets such as SkM$^*$\cite{SkMS}
are known to provide a reasonable
description of nuclear deformations.
To check sensitivity to the force parameters 
we performed a  calculation using SkM$^*$.
The calculated potential energy curves associated with 
an oblate HF solution in \Se\ and a spherical HF solution in \Zr\
are shown in Fig.4 (a) and (c). Although they do not have a minima
at finite $\alpha_{3m}$, the potential curve shows softness 
for the tetrahedral $\alpha_{32}$ direction of the \Zr\
spherical state ($\beta=0$) and for the triangular $\alpha_{33}$
direction of the \Se\ oblate state. This  qualitative agreement
with the SIII results   can be expected since 
the single-particle structures which play a central role for 
the mechanism for the non-axial octupole deformations are not very
different for both parameter sets. In Fig.4(b) and (d), we show
the results of SIII but with increased pairing force
strength  in order to show the
sensitivity to the pairing correlation. 
Here we assumed 10\% increase of $G_{n,p}$ which is within a
reasonable range reproducing the experimental pairing gap (the odd-even mass
difference).
From comparison
with Fig.2, it is seen that the pairing correlation have an effect to
reduce the non-axial deformation. Note, however, that the essential
features of the potential curve remain the same.

\subsection*{Spectral signatures}

If a rigid tetrahedron deformation is realized, a characteristic
pattern of excited spectra is expected. 
Rotational excitation of a even-even tetrahedron nucleus  will accompany 
a sequence of levels $0^+, 3^-, 4^+, 6^+,7^-,...$ following
the rotational energy relation $E(I)-E(0)=I(I+1)/2\cal{J}$ 
in parallel with the tetrahedron molecule \cite{molecule}.
Here the spectrum is labeled by the irreducible representation
of the \Td\ group.
The rotational levels are connected
by strong E3 transitions instead of E2. Viewing the potential energy
curve for the tetrahedral solution in \Zr, we can expect not a 
rigid deformation but rather a transitional situation between the
tetrahedron rotation and the octupole vibration around 
the spherical state.
From the experimental systematics of the first 3$^-$ excitation 
energy in the neighboring mass region \cite{Cottle}, 
a typical octupole vibrational energy is about a few MeV.
Using this estimate and the calculated potential energy curve for the 
$\alpha_{32}$ direction, amplitude associated with the transitional
tetrahedral excitation is estimated to be about $\alpha_{32}(\beta_3) \sim
0.3$.  This corresponds to $B({\rm E3}) \sim 60$ W.u. on the basis of
$B({\rm E3}) =(3ZeR^3\beta_3/4\pi)^2$, which is much larger than
the systematic values \cite{BE3} in neighboring mass region.

For the oblate state in \Se, the potential energy curve suggests
soft vibrational excitation for the \Ythree\ direction which couples
with the rotational excitation. 
We can expect for such case that the excitation spectra showing
a $K^\pi=0^+$ rotational band ($0^+,2^+,4^+,..$) 
associated with the ground state, and 
a $K^\pi=3^-$ rotational band ($3^-,4^-,5^-,..$) 
associated with the vibrational excitation. 
From the calculated potential energy curve, the 
vibrational amplitude is estimated as $\alpha_{33}(\beta_3) \sim 0.2-0.3$,
corresponding to $B({\rm E3}) \sim 20-40$ W.u. for interband
transitions.

The nuclei under discussion are close to the proton drip line,
and experimental information on the excited spectra is not very rich
at present.
Candidates of negative parity states are found in \Ge, which suggests presence
of some but not very strong octupole collectivity\cite{GeEnnis}. 
Quite recently candidate
negative parity states are found in \Se, and the observed
spectra looks similar to that in \Ge\ although the experimental
information is not enough to see the octupole collectivity
\cite{SeSkoda}.

\section{Conclusions}

The fully three dimensional Hatree-Fock calculation suggests that the
non-axial octupole deformations are important in $N=Z, A\sim60-80$
nuclei. Prominent examples are the tetrahedral deformation of
an shape coexisting excited state in \Zr, and the triangular 
softness of the oblate state in \Se. 
The tetrahedron deformation causes a shell gap at $N,Z=40$, which 
founds the microscopic origin of the exotic shape.

\end{document}